# COMPARATIVE ALGORITHMIC GOVERNANCE OF PUBLIC HEALTH INSTRUMENTS ACROSS INDIA, EU, US AND LMICs

SAHIBPREET SINGH[*]

**Abstract**

*This study investigates the juridico-technological architecture of international public health instruments. It focuses specifically on their differential implementation across India, the European Union, the United States, and LMICs, particularly in Sub-Saharan Africa. Despite the proliferation of binding nature the International Health Regulations (IHR 2005) and the WHO FCTC, jurisdictional asymmetries in compliance remain pronounced. The study addresses a critical research lacuna: the inadequate harmonisation between normative health law and algorithmic public health infrastructures in resource-constrained and technologically underserved jurisdictions. The principal objective is to assess the extent to which artificial intelligence augments IPHI implementation while analysing emergent doctrinal and infrastructural bottlenecks. Employing a comparative doctrinal analysis and legal-normative mapping, the study triangulates legislative instruments, WHO monitoring frameworks, AI applications (e.g., BlueDot, Aarogya Setu, EIOS), and compliance metrics. Preliminary results demonstrate that while AI technologies have enhanced early detection, surveillance precision, and health system responsiveness in high-capacity jurisdictions, LMICs continue to encounter impediments due to infrastructural scarcities, data privacy vacuums, and fragmented legal scaffolding. Notably, the EU's Artificial Intelligence Act and GDPR present promising regulatory prototypes for health-oriented algorithmic governance, whereas LMICs exhibit embryonic AI integration, limited internet penetration, and insufficient jurisprudential coherence. The findings underscore the imperative of embedding AI within a rights-compliant, supranationally coordinated regulatory framework to ensure equitable health outcomes and compliance with global normative instruments. This research advances the discourse by proposing a model for global algorithmic treaty-making inspired by FCTC architecture and advocates for WHO-led compliance mechanisms modelled on WTO's Dispute Settlement Body. The study offers a prescriptive framework aimed at reconciling technological modernisation with public health legality to fortify pandemic preparedness, surveillance equity, and transnational governance resilience.*

---

[*] PhD Research Scholar, School of Law, Lovely Professional University, Phagwara, Punjab.

## 1. Introduction

The confluence of jurisprudence and medical science has crystallised into an indispensable domain for confronting transnational health issues. IPHIs[1] constitute the normative architecture for harmonising legal systems. They also serve as foundational in harmonising healthcare systems. WHO defines health as a fundamental right.[2] FCTC[3] targets non-communicable diseases by regulating tobacco use.[4] These instruments articulate binding and non-binding obligations designed to advance health equity. This chapter examines how nations apply these instruments. India's constitutional right to health under Art. 21 drives its public health policies.[5] The EU coordinates health governance via the Cross-Border Healthcare Directive (2011/24/EU).[6] US relies on the PHSA[7] within a decentralised framework.[8] LMICs often face resource constraints that limit compliance.[9] A comparative analysis reveals best practices in achieving global health objectives. AI on other hand transforms public health governance by enhancing instrument implementation. Predictive AI models enable early disease detection, aligning with IHR[10] requirements.[11] e.g., BlueDot's early detection of COVID-19; India's Aarogya Setu app demonstrates AI-driven contact tracing during COVID-19.[12] By integrating AI responsibly, nations can strengthen compliance while upholding health equity.[13] The analysis spans India, EU, US, alongside LMICs to address emerging technological opportunities.

---

[1] International Public Health Instruments.
[2] Lawrence O. Gostin and Devi Sridhar, "Global Health and the Law," 370 *New England Journal of Medicine* 1732–40 (2014).
[3] Framework Convention on Tobacco Control.
[4] Hadii M. Mamudu and Donley T. Studlar, "Multilevel Governance and Shared Sovereignty: European Union, Member States, and the FCTC," 22 *Governance* 73–97 (2009).
[5] Sushant Chandra and Shireen Moti, "Jurisprudential Reimagination on Rights to Education and Healthcare in India: In Pursuit of a Coherent Theory," 46 *Liverpool Law Review* 85–99 (2025).
[6] Takis Vidalis and Irini Kyriakaki, "Cross-border Healthcare: Directive 2011/24 and the Greek Law," 21 *European Journal of Health Law* 33–45 (2014).
[7] Public Health Service Act, 42 U.S.C. § 201 et seq. (1944) (U.S.).
[8] J. Rogers Hollingsworth and Robert Hanneman, *Centralization and Power in Social Service Delivery Systems* (Springer Netherlands, Dordrecht, 1984), III.
[9] GRID COVID-19 Study Group, "Combating the COVID-19 pandemic in a resource-constrained setting: insights from initial response in India," 5 *BMJ Global Health* e003416 (2020).
[10] International Health Regulations 2005.
[11] Abhishek Tripathi and Rachna Rathore, "AI in Disease Surveillance — An Overview of How AI Can Be Used in Disease Surveillance and Outbreak Detection in Real-World Scenarios," 1st ed., in R. Singh, A. Gehlot, *et al.* (eds.), *AI in Disease Detection* 337–59 (Wiley, 2025).
[12] Mandira Narain, "Practices on Aarogya Setu: Mapping Citizen Interaction with the Contact-Tracing App in the Time of COVID-19," 18 *East Asian Science, Technology and Society: An International Journal* 334–54 (2024).
[13] Raj M. Ratwani, Karey Sutton and Jessica E. Galarraga, "Addressing AI Algorithmic Bias in Health Care," 332 *JAMA* 1051 (2024).

## 2. International Public Health Instruments: Frameworks and Constitutive Principles

### 2.1 Synoptic Overview

International instruments operate as the constitutive superstructure of transnational health governance, serving as the juridico-normative substratum that undergirds cooperative engagement among sovereign actors. These instruments embody the structural nucleus to regulate transnational health threats, plus operationalise equality in the administration of therapeutic interventions. These further instantiate uniformity in the dispensation of remedial modalities. The WHO's Constitution (1948) is a foundational document. It codifies health as an integrative state of complete corporeal, psychological, and societal well-being, conceived in its fullest plenitude. It does not merely signify the absence of disease or infirmity.[14] Its Art. 1 obligates WHO to elevate universal health parameters, mandating to advance the highest attainable standard of health. This establishes a human rights oriented framework which exerts influence upon successive normative instruments.[15] This doctrinal principle permeates the UDHR[16] under Art. 25.[17] It codifies health as an inalienable entitlement inherent to the human condition. This right likewise attains juridico-normative recognition, as enunciated in Art. 12.[18] (within ICESCR[19]). IHR form an indispensable linchpin within the collective health security framework. It imposes binding obligations upon 196 state entities including India. These regulations obligate the prevention of PHEICs,[20] necessitating the establishment of disease surveillance and expeditious response apparatuses.[21] They necessitate the state entities to notify WHO regarding potential PHEICs within a temporal threshold of twenty-four hours.[22] The proclamation of the COVID-19 PHEIC in Jan. 2020 exemplified the instrumental role of the IHR in orchestrating transnational countermeasures.[23] These regulatory provisions govern restrictions upon transnational mobility, governing port-of-entry sanitary protocols. They harmonise health security imperatives with obligations under international trade regimes. The FCTC of WHO represents the first treaty in the domain of global public health. It attained ratification by 182 sovereign entities including India.[24] It imposes binding normative obligations upon state parties, necessitating the diminution of tobacco consumption. It mandates the inscription of graphic cautionary insignia as health warnings upon commercial packaging to attenuate consumption patterns. These obligations mandate the institutionalisation of smoke-free environments.[25] The Framework Convention on Tobacco Control embodies a calibrated regulatory stratagem directed towards the containment of non-communicable diseases. These pathological conditions account for seventy-one percent of the global mortality index per annum.[26] The Sustainable Development Goal 3, another instrument, further integrates health imperatives into the schema of global developmental trajectories. It put emphasis on the attainment of UHC[27].[28]

---

[14] David Matthew Doyle and Bruce G. Link, "On social health: history, conceptualization, and population patterning," 18 *Health Psychology Review* 626–55 (2024).

[15] Janet E. Lord, David Suozzi and Allyn L. Taylor, "Lessons from the Experience of U.N. Convention on the Rights of Persons with Disabilities: Addressing the Democratic Deficit in Global Health Governance," 38 *Journal of Law, Medicine & Ethics* 564–79 (2010).

[16] Universal Declaration of Human Rights.

[17] Christine Chinkin, "Health and Human Rights," 120 *Public Health* 52–9 (2006).

[18] *Ibid.*

[19] International Covenant on Economic, Social and Cultural Rights.

[20] Public Health Emergency of International Concern.

[21] Lawrence O. Gostin and Rebecca Katz, "The International Health Regulations: The Governing Framework for Global Health Security: The International Health Regulations," 94 *The Milbank Quarterly* 264–313 (2016).

[22] Youngmee Jee, "WHO International Health Regulations Emergency Committee for the COVID-19 outbreak," 42 *Epidemiology and Health* e2020013 (2020).

[23] Roojin Habibi et al., "Do not violate the International Health Regulations during the COVID-19 outbreak," 395 *The Lancet* 664–6 (2020).

## 2.2 Legal Principles and Obligations

The IPHIs rest on the legal principles balancing state autonomy with global cooperation. The doctrinal tenet of state sovereignty vests polities with the prerogative to modulate IHR obligations so as to facilitate concerted action against global health threats.[29] However, Art. 43 of IHR obliges sovereign entities to adduce credible scientific vindication as a *sine qua non* for implementation. Consequently, this obligation forestalls gratuitous perturbations in cross-border commercial intercourse.[30] The human rights-anchored model accords pre-eminence to the equitable dispensation of healthcare.[31] It drives WHO's Essential Medicines List policy which aspires to institutionalise the availability of pharmacology at economically viable thresholds.[32] Universal Health Coverage initiatives exemplify this operationalisation. Legal enforceability exhibits considerable variation across instruments. IHR & FCTC constitute a binding juridical corpus that impose obligatory commitments upon state entities. Their compliance remain subject to monitoring through reporting mechanisms instituted by the WHO. WHO Constitution and SDGs retains the character of *lex mitis*, thereby operating within the realm of soft law. These soft law instruments derive normative force from moral or political imperatives.[33] Its operational efficacy remains contingent upon the vicissitudes of political concordance. These binding instruments encounter formidable impediments in enforcement praxis. These difficulties emanate from the paucity of supranational sanctions. The latency in compliance discerned during the Ebola contagion of 2014 exemplifies this structural limitation.[34] The doctrine of proportionality presides over the imposition of restrictive public health measures. These measures include the interdiction of trans-boundary mobility. They necessitate a calibrated equilibrium between collective sanitary and safeguarding of inviolable individual liberties. This principle harmonises with the Siracusa Principles on the Limitation and Derogation of Provisions embedded within ICCPR.[35]

---

[24] Sonu Goel et al., "Countering Tobacco Industry Interference: A Policy Brief on Strengthening the WHO FCTC Article 5.3 Adoption in India," 49 *Indian Journal of Community Medicine* S228–33 (2024).
[25] Rob Cunningham, "Tobacco package health warnings: a global success story," 31 *Tobacco Control* 272–83 (2022).
[26] Douglas Bettcher et al., "Tobacco as Global Health Risk Factor: Disease Burden, Preventive Action and Regulatory Challenges," in R. Haring, I. Kickbusch, *et al.* (eds.), *Handbook of Global Health* 1–62 (Springer International Publishing, Cham, 2021).
[27] Universal Health Coverage.
[28] Robert Kokou Dowou et al., "Increased investment in Universal Health Coverage in Sub–Saharan Africa is crucial to attain the Sustainable Development Goal 3 targets on maternal and child health," 81 *Archives of Public Health* 34 (2023).
[29] Rana Sulieman et al., "International legal issues of national sovereignty and authority impacting global health security" *Modernizing Global Health Security to Prevent, Detect, and Respond* 71–85 (Elsevier, 2024).
[30] Roojin Habibi et al., "Do not violate the International Health Regulations during the COVID-19 outbreak," 395 *The Lancet* 664–6 (2020).
[31] Lawrence O. Gostin and Devi Sridhar, "Global Health and the Law," 370 *New England Journal of Medicine* 1732–40 (2014).
[32] Veronika J Wirtz et al., "Essential medicines for universal health coverage," 389 *The Lancet* 403–76 (2017).

### 2.3 Challenges in Implementation

The operationalisation of international public health instruments encounters formidable impediments. These primarily concern the reconciliation of universalised standards with the realities of domestic jurisdictions. Jurisdictional dissonance materialises when municipal legislations deviate from transnational obligations. EDA 1897[36] exemplifies such deviation. It remains devoid of advanced surveillance mechanisms mandated under IHR. This legislative inadequacy attenuates its efficacy in the containment of pandemic contingencies. It acts as a lacuna and obstructs compliance with IHR.[37] Resource incapacities in LMICs further obstructs to its adherence. These incapacities materialise in the form of deficient medical infrastructures, insufficiently trained healthcare workforce. Empirical assessments reveal that merely thirty percent of African countries fulfilled these stipulated requisites by 2023.[38] Emergent concerns further complicate the operationalisation. Anthropogenic climatic changes serve as catalytic agents for the exponential proliferation of pathogenic vectors which necessitate the evolution of juridical structure capable of adapting regulatory responses.[39] Malaria exemplifies this pathogenic proliferation as a disease vector. Antimicrobial resistance on the other hand constitutes a parallel menace that takes an annual mortality toll of approximately 1.27 million human lives. This existential threat mandates coordinated enforcement which transcends the regulatory ambit of these extant legal instruments.[40] The COVID-19 pandemic further unveiled the fragility of IHR implementations in which deferred warnings and inconsistent imposition of travel restrictions vitiated the efficacy of global pandemic responses and public health security measures.[41] Technological inconsistencies intensify these systemic inequities within global health governance. AI surveillance augments epidemiological capacities within prosperous countries whereas LMICs remain encumbered by infra & tech incapacities. This technological lacuna creates a deep-seated digital bifurcation.[42] Juridical scaffolding for AI in healthcare in LMICs remains conspicuously deficient in multiple legal orders in contrast to EU's AI Act that embodies a paradigmatic demonstration of governance regime.[43] This vacuum imperils the equitable operationalisation of data-intensive health interventions. These challenges accentuate the need for doctrinal and technological harmonisation encompassing the transformative capacity of artificial intelligence within the domain of international public health governance.[44]

---

[33] Adnan Mahmutovic and Abdulaziz Alhamoudi, "Understanding the Relationship between the Rule of Law and Sustainable Development," 7 *Access to Justice in Eastern Europe* 170–97 (2023).
[34] Suerie Moon et al., "Will Ebola change the game? Ten essential reforms before the next pandemic. The report of the Harvard-LSHTM Independent Panel on the Global Response to Ebola," 386 *The Lancet* 2204–21 (2015).
[35] Adetoun Adebanjo and Ebenezer Durojaye, "International Human Rights Norms and Standards on Derogation and Limitation of Rights During a Public Emergency," in E. Durojaye, D. M. Powell (eds.), *Constitutional Resilience and the COVID-19 Pandemic* 79–109 (Springer International Publishing, Cham, 2022).
[36] Epidemic Diseases Act, No. 3 of 1897 (India).
[37] Amita Singh, "Covid-19 Pandemic and the Future of SDGs," in V. K. Malhotra, R. L. S. Fernando, *et al.* (eds.), *Disaster Management for 2030 Agenda of the SDG* 279–317 (Springer Singapore, Singapore, 2020).
[38] Greta Cranston, "Understanding the expectations, positions and ambitions of LMICs during pandemic treaty negotiations, and the factors contributing to them," 5, in E. Okereke (ed.), *PLOS Global Public Health* e0003851 (2025).
[39] Sadie J. Ryan, Catherine A. Lippi and Fernanda Zermoglio, "Shifting transmission risk for malaria in Africa with climate change: a framework for planning and intervention," 19 *Malaria Journal* 170 (2020).

## 3. Comparative Perspectives: Implementation Across Jurisdictions

### 3.1 India: Public Health Law and Policy

Health policy derives its normative foundation from the constitutional enshrinement of the right to health. This entitlement subsists implicitly within Art. 21 of the Constitution of India. The aforementioned provision affirms the right to life and individual liberty. Judicial interpretations, as demonstrated in *PBKMS v. State of WB*[45], have augmented its scope.[46] This construction has incorporated elevated access to timely medical intervention into the realm of fundamental rights.[47] Statutory enactments fortify this constitutional commitment. The EDA constitutes a principal legislation (amended in 2020). It confers plenary authority upon the governmental apparatus to impose restrictive measures. Such measures include compulsory lockdowns designed to suppress the transmission of infectious diseases.[48] Despite such augmentation, the Act remains structurally anachronistic. Its antiquated provisions preclude the institution of broad monitoring apparatus. This statutory insufficiency impedes consonance with the IHR, which mandates the robust disease detection and obligatory notification systems.[49] NHM[50] aspires to fortify the healthcare set-up, prioritising operations in non-urban territorial units.[51] Its efficacy remains circumscribed by fiscal insufficiency and asymmetric implementation across states. The CEA[52] seeks to regularise private healthcare dispensaries. The Act prescribes infrastructural and clinical service-oriented parameters. The corpus of compliance remains limited. As of 2023, only eleven states have operationalised full statutory adoption.[53] The response to the 2019 pandemic exemplifies the country's efforts towards adherence with the stipulations of IHR. The government apparatus instituted an unprecedented cessation of civil mobility as nationwide lockdowns. The Integrated Disease Surveillance Programme was operationalised to facilitate real-time monitoring.[54] Despite all this, the operational lacunae emerged. The insufficiency of diagnostic infrastructure further illuminated lacunae within India's core capacities mandated under IHR.[55] The deployment of the Aarogya Setu application represented an unprecedented advancement in epidemiological surveillance. The application incorporated AI functionalities to facilitate contact tracing procedures, however legal apprehensions surfaced. It's functionality provoked critical scrutiny regarding data privacy. These apprehensions pertained to statutory provisions codified within the IT Act[56]. Specific concerns emerged regarding data retention and user consent.[57] Seventy percent of healthcare expenditure remains financed through out-of-pocket disbursements by the populace. A pronounced disparity subsists between urban and non-urban metrics. Urban territories exhibit a contrasting density of 16.1 medical practitioners per equivalent population strata compared to a mere 3.2 licensed practitioners per 10,000 inhabitants in non-urban regions.[58] These systemic deficiencies collectively circumscribe India's capacity to effectuate comprehensive operationalisation of global public health instruments. Remedial recalibration necessitates comprehensive legal reconfiguration and technological modernisation.

---

[40] Muhammad Usman Qamar and Aatika, "Impact of Climate Change on Antimicrobial Resistance Dynamics: An Emerging One Health Challenge," 18 *Future Microbiology* 535–9 (2023).
[41] Adam Ferhani and Simon Rushton, "The International Health Regulations, COVID-19, and bordering practices: Who gets in, what gets out, and who gets rescued?," 41 *Contemporary Security Policy* 458–77 (2020).
[42] Muhammad Salar Khan, Hamza Umer and Farhana Faruqe, "Artificial intelligence for low income countries," 11 *Humanities and Social Sciences Communications* 1422 (2024).
[43] Vishambhar Raghuwanshi, Pranjal Khare and Paridhi Sharma, "Global AI Governance and Collaboration: The Role of International Bodies in AI Regulation," in A. Youssef, A. Arslan (eds.), *Advances in Computational Intelligence and Robotics* 225–50 (IGI Global, 2025).
[44] Mark L Flear, "Expectations as techniques of legitimation? Imagined futures through global bioethics standards for health research," 8 *Journal of Law and the Biosciences* lsaa086 (2021).
[45] *Paschim Banga Khet Mazdoor Samity v. State of W.B.*, (1996) 4 SCC 37 (India).
[46] K Mathiharan, "The fundamental right to health care," 11 *Indian Journal of Medical Ethics [Online]* 123 (2016).
[47] Kiran Kumar Gowd, Donthagani Veerababu and Veeraiahgari Revanth Reddy, "COVID -19 and the legislative response in India: The need for a comprehensive health care law," 21 *Journal of Public Affairs* e2669 (2021).

---

[48] Kritika Maheshwari, "Responding to Covid-19 in India: Reducing Risk or Increasing Domination?," in P. R. Brown, J. O. Zinn (eds.), *Covid-19 and the Sociology of Risk and Uncertainty* 29–51 (Springer International Publishing, Cham, 2022).
[49] Manish Tewari, "India's Fight Against Health Emergencies: In Search of a Legal Architecture" *ORF Issue Brief* (2020).
[50] National Health Mission.
[51] Aakriti Grover and R. B. Singh, "Health Policy, Programmes and Initiatives" *Urban Health and Wellbeing* 251–66 (Springer Singapore, Singapore, 2020).
[52] Clinical Establishments (Registration and Regulation) Act, 2010, Act No. 23 of 2010 (India).
[53] Dipika Jain, "Regulation of Digital Healthcare in India: Ethical and Legal Challenges," 11 *Healthcare* 911 (2023).
[54] Manish Tewari, "India's Fight Against Health Emergencies: In Search of a Legal Architecture" *ORF Issue Brief* (2020).
[55] Arista Lahiri et al., "Effectiveness of preventive measures against COVID-19: A systematic review of In Silico modeling studies in Indian context," 64 *Indian Journal of Public Health* 156 (2020).
[56] Information Technology Act, No. 21 of 2000 (India).
[57] Dipika Jain, "Regulation of Digital Healthcare in India: Ethical and Legal Challenges," 11 *Healthcare* 911 (2023).

### 3.2 European Union: Harmonised Health Governance

EU acts as a supranational body, standardising governance. The framework is entrenched within a coherent health governance taxonomy. This institutional architecture aspires towards the calibration of equilibrium between the sovereign entities of constituent national polities. TF-EU[59] encapsulates this norm within the ambit of Art. 168. This provision empowers the Union to furnish auxiliary support to public health policy formulation. It simultaneously enjoins deference to the reserved competences of national jurisdictions.[60] The Cross-Border Healthcare Directive enacted under legislative instrument 2011/24/EU institutionalises patient mobility across member states. It prescribes standardised quality parameters for healthcare. This legal regime remains consonant with the WHO's articulated postulates pertaining to the right to health.[61] The ECDC[62] occupies a pivotal institutional locus within the matrix of IHR compliance. The Centre exercises coordination authority over disease surveillance mechanisms, orchestrating coordinated responses to transnational contingencies including pandemics.[63] The 2019 crisis epitomised this functional mandate. The Early Warning and Response System of the European Centre for Disease Prevention and Control facilitated expeditious informational dissemination. Notwithstanding the inconsistent policy measures undertaken by discrete member states, functional asymmetries materialised. The delayed imposition of territorial lockdown protocols within the Italian jurisdiction embodied this discordance. Latent tensions are thereby revealed within the European layered system.[64] The GDPR[65] demarcates prescriptive and rigorous regulatory standards delineating the custodial obligations pertaining to health-related data. This regulatory regime exerts direct implications upon artificial intelligence driven public health mechanisms. This includes digital constructs designed for contact tracing.[66] The Artificial Intelligence Act designates health-oriented AI systems as constituting a high-risk operational category. The Act mandates procedural transparency, also imposing accountability strictures congruent with the ethical surveillance.[67] The European Union exhibits institutional fortitude. This robustness derives from the confluence of expansive legal codifications and substantial fiscal endowments. EU4Health Programme exemplifies this fiscal commitment (2021–27). It remains an archetypal expression of budgetary prioritisation. The Programme allocates a monetary corpus quantified at 5.3 billion Euros towards health system resilience.[68] Notwithstanding these systemic strengths, latent challenges subsist, including Brexit's disruption. The United Kingdom's secessionary trajectory from the European Union disrupted pre-existing health governance synchronisations between the British and Continental jurisdictions.[69] Variability in infrastructural capacity afflicts member polities. This disjunction remains especially pronounced within the Eastern European territorial bloc.[70] The harmonised juridical and administrative apparatus of the EU yields instructive prototype for global health governance. Nevertheless, the internal dissonances and external misalignments with non-EU normative regimes delimit its integrative efficacy.[71]

---

[58] Megha Kapoor et al., "Impact of COVID-19 on healthcare system in India: A systematic review," 12 *Journal of Public Health Research* 22799036231186349 (2023).
[59] The Treaty on the Functioning of the European Union.
[60] Oliver Bartlett and Anja Naumann, "Reinterpreting the health in all policies obligation in Article 168 TFEU: the first step towards making enforcement a realistic prospect," 16 *Health Economics, Policy and Law* 8–22 (2021).
[61] Alceste Santuari, "The European Union Directive on the application of patients' rights in cross-border healthcare. Could it be part of the Global Health Summit strategy?," 33, in C. Bottari (ed.), *International Journal of Risk & Safety in Medicine* 125–31 (2022).
[62] European Centre for Disease Prevention and Control.
[63] Anna Durrance-Bagale et al., "Operationalising Regional Cooperation for Infectious Disease Control: A Scoping Review of Regional Disease Control Bodies and Networks" *International Journal of Health Policy and Management* 1 (2021).
[64] Sarah Cataldi et al., "Monitoring and Early Warning System: Regional Monitoring Strategy in Lombardy Region," 6 *Epidemiologia* 7 (2025).
[65] General Data Protection Regulation.

### 3.3 United States: Decentralised Health Governance

The United States operationalises a decentralised architecture of public health governance. This constitutional architecture demarcates authority *inter se* across the federal tier and its constituent subnational sovereignties. PHS Act[72] confers institutional prerogatives upon CDC[73]. This federal agency assumes the mandate to orchestrate compliance with IHR. Its operational remit encompasses disease surveillance, with its procedural functions including the imposition of quarantine measures.[74] The autonomous authority exercised by individual state entities engenders inconsistent regulatory execution.[75] This phenomenon materialised conspicuously during the 2019 pandemic. NY[76] State enforced stringent lockdown protocols. In contradistinction, State of Florida privileged economic activity resumption.[77] HIPAA[78] constitutes the principal statutory instrument governing the privacy of health-related data. Its regulatory scope remains markedly circumscribed when juxtaposed with the European Union's General Data Protection Regulation. This regulatory lacuna engenders complexities in the operationalisation of algorithmic public health technologies.[79] The opioid epidemic constitutes the *sine qua non* for regulatory recalibration. It further necessitates targeted juridico-normative interventionism within the domain of public health. This epidemic precipitated in excess of seventy thousand fatalities attributable to overdose phenomena during the year 2021.[80] CSA[81] prescribes regulatory canons for narcotic dispensation. State-specific prescription monitoring regimes complement this corpus by endeavouring to attenuate pharmaceutical misuse. However, the disarticulated nature of enforcement circumscribes systemic efficacy.[82] Artificial Intelligence-driven applications incorporating predictive analytics for overdose susceptibility operate under the regulatory auspices of the FDA[83]. This federal entity categorises algorithmic tools under the regulatory apparatus articulated in Title 21 of the CFR[84] Part 860.[85] US exhibits considerable infrastructural strengths within the technological sphere. Its private sector further engenders substantial innovation within health informatics.[86] On the other hand, it also endures institutional incapacities within the healthcare systemic matrix. Frictions between federal and state authority create regulatory disharmony. With regard to distributive asymmetries in healthcare accessibility, the demographic data indicate that 8.6 percent of the national population subsisted without health insurance coverage during 2023.[87] This decentralised constitutional configuration of the United States exhibits profound deviation. By contrast, EU institutionalises a homogenised governance corpus whose normative construct, operating as a *lex specialis*, is enshrined within its *acquis communautaire* and supranational regulatory taxonomy. This structural bifurcation accentuates the jurisprudential incongruity between regulatory adaptability and rigidity in consonance of IHR.[88]

---

[66] Nikolaus Forgó et al., "Big Data, AI and health data: between national, European, and international legal frameworks," in A. Zwitter, O. Gstrein (eds.), *Handbook on the Politics and Governance of Big Data and Artificial Intelligence* 358–94 (Edward Elgar Publishing, 2023).

[67] Sofia Palmieri, "The Renewed EU Legal Framework for Medical AI," 15 *European Journal of Law and Technology* 23 (2024).

[68] Dimitrios G. Katehakis et al., "The smartHEALTH European Digital Innovation Hub experiences and challenges for accelerating the transformation of public and private organizations within the innovation ecosystem," 11 *Frontiers in Medicine* 1503235 (2024).

[69] Mark Dayan et al., "Parallel, divergent or drifting? Regulating healthcare products in a post-Brexit UK," 30 *Journal of European Public Policy* 2540–72 (2023).

[70] Sabina Stan and Roland Erne, "Time for a paradigm change? Incorporating transnational processes into the analysis of the emerging European health-care system," 27 *Transfer: European Review of Labour and Research* 289–302 (2021).

[71] Susan Bergner, "The role of the European Union in global health: The EU's self-perception(s) within the COVID-19 pandemic," 127 *Health Policy* 5–11 (2023).

[72] Public Health Service Act, 42 U.S.C. § 201 et seq. (1944) (U.S.).

[73] Centers for Disease Control and Prevention.

[74] Antony J. Blinken and Xavier Becerra, "Strengthening Global Health Security and Reforming the International Health Regulations: Making the World Safer From Future Pandemics," 326 *JAMA* 1255 (2021).

---

[75] Katie Hogan et al., "Contact Tracing Apps: Lessons Learned on Privacy, Autonomy, and the Need for Detailed and Thoughtful Implementation," 9 *JMIR Medical Informatics* e27449 (2021).
[76] New York.
[77] Arianna Vedaschi, "General Report: Governmental Policies to Fight Pandemic: The Boundaries of Legitimate Limitations on Fundamental Freedoms," in A. Vedaschi (ed.), *Governmental Policies to Fight Pandemic* 3–105 (Brill | Nijhoff, 2024).
[78] Health Insurance Portability and Accountability Act (HIPAA), 42 U.S.C. § 1320d et seq. (1996) (U.S.).
[79] Leslie Lenert and Brooke Yeager McSwain, "Balancing health privacy, health information exchange, and research in the context of the COVID-19 pandemic," 27 *Journal of the American Medical Informatics Association* 963–6 (2020).
[80] Foojan Zeine et al., "Solving the Global Opioid Crisis: Incorporating Genetic Addiction Risk Assessment with Personalized Dopaminergic Homeostatic Therapy and Awareness Integration Therapy," 8 *Journal of Addiction Psychiatry* 50–95 (2024).
[81] Controlled Substances Act, 21 U.S.C. § 801 et seq. (1970) (U.S.).
[82] Bhanu Prakash Kolla et al., "Prescribing controlled substances in sleep medicine clinics: an overview of legal issues and best safety practices" *Journal of Clinical Sleep Medicine* jcsm.11770 (2025).
[83] Food and Drug Administration.
[84] Code of Federal Regulations.
[85] S Shamtej Singh Rana, Jacob S. Ghahremani and Ronald A. Navarro, "The future of Food and Drug Administration regulation on artificial intelligence–enabled medical devices: an orthopedic surgeon's guide," 34 *Journal of Shoulder and Elbow Surgery* 260–4 (2025).
[86] Eric J. Topol, "High-performance medicine: the convergence of human and artificial intelligence," 25 *Nature Medicine* 44–56 (2019).

### 3.4 LMICs: Case Study (Sub-Saharan Africa)

LMICs, located mainly within the Sub-Saharan African territorial constituencies endure entrenched infrastructural and institutional obstructions in the operationalisation of international public health instruments. FCTC issued under the aegis of the WHO exhibits distinct enforcement deficiencies. Its enforcement apparatus remains tenuous within territorial entities, such as Nigeria. Merely seventeen percent of commercial tobacco packaging complies with the prescriptive warning label stipulations mandating health warnings. The underlying cause resides in the fragility of domestic regulatory architectures.[89] IHR imposes core capacity prerequisites including laboratory infrastructure. These stipulations remain unfulfilled by seventy percent of African nations. This deficiency is exacerbated by an entrenched dependence on extrinsic financial endowments. In multiple sovereign jurisdictions, close to fifty percent of health-sector budgetary disbursements are underwritten by donor mechanisms.[90] The institutional architecture of the Africa CDC is mandated to reinforce juridico-regulatory alignment with IHR but its functional capacity remains delimited by persistent resource insufficiency. The delayed response to the 2014–16 Ebola outbreaks highlights underlying structural failures.[91] Artificial intelligence deployment within low and middle-income polities remains embryonic. Initiatives integrating mobile-mediated epidemiological surveillance in Kenya evince nascent potentiality.[92] These operational constraints demonstrate potential efficacy while simultaneously encountering formidable impediments. The principal constraint resides in the attenuated accessibility of digital infrastructure. Internet penetration as of 2023 indices continue to exhibit marked attenuation at 29% in Sub-Sahara.[93] Prescriptive juridico-regulatory frameworks, operating as a *lex specialis*, codify the governance taxonomy of health data and AI operationalisation which remain conspicuously non-existent within these territorial jurisdictions. This lacuna provokes augmented susceptibility to infringements of informational privacy.[94] Practical indices derived from LMICs underscore the *sine qua non* for the systematic augmentation of institutional capacity taxonomies. Technical assistance initiatives as administered by WHO in response to structural insufficiencies as well coalition by the Global Alliance for Vaccines and Immunisation have substantively enhanced vaccine dissemination indices.[95] These exemplars illuminate the entrenched asymmetries inherent in the system. These necessitate the articulation of jurisdiction-specific laws. Such interventions are crucial to facilitate the congruence of LMICs with prevailing trans-national normative standards.

---

[87] Robin A. Cohen and Michael E. Martinez, *Health Insurance Coverage: Early Release of Estimates From the National Health Interview Survey, January–June 2023* 22 (National Center for Health Statistics, USA, 2023).

[88] Thorbjørn Sejr Guul, Anders R. Villadsen and Jesper N. Wulff, "Does Good Performance Reduce Bad Behavior? Antecedents of Ethnic Employment Discrimination in Public Organizations," 79 *Public Administration Review* 666–74 (2019).

[89] Adetoun T. Adebanjo, "5 An overview of Nigeria's efforts in addressing non-communicable diseases within the WHO FCTC" *International Human Rights Law and the Framework Convention on Tobacco Control: Lessons from Africa and Beyond* (Taylor & Francis, 2022).

[90] Anne Doble et al., "The role of international support programmes in global health security capacity building: A scoping review," 3, in C. J. Standley (ed.), *PLOS Global Public Health* e0001763 (2023).

[91] Chulwoo Park, "Lessons learned from the World Health Organization's late initial response to the 2014-2016 Ebola outbreak in West Africa," 13 *Journal of Public Health in Africa* (2022).

[92] Abayomi O. Agbeyangi and Jose M. Lukose, "Telemedicine Adoption and Prospects in Sub-Sahara Africa: A Systematic Review with a Focus on South Africa, Kenya, and Nigeria," 13 *Healthcare* 762 (2025).

| DOMINION. | KEY FRAMEWORKS. | IHR 2005 COMPLIANCE. | FCTC IMPLEMENTATION. | TECH INTEGRATION. | KEY ISSUES. |
|---|---|---|---|---|---|
| INDIA | i. Constitution, Art. 21; ii. EDA, 1897 (amended 2020); iii. CEA, 2010 | a) Partial: IDSP established, but delayed COVID-19 reporting; b) Weak surveillance infrastructure | – Partial: Tobacco warning labels, weak rural enforcement | [1] Aarogya Setu app (AI-based contact tracing) [2] Data privacy concerns under IT Act, 2000 | a. Fragmented healthcare system; b. Rural-urban disparities; c. NHM underfunding |
| EUROPEAN UNION | i. TFEU, Art. 168; ii. CBHD (2011/24/EU); iii. GDPR (2016/679) | a) Strong: ECDC coordinates via EWRS; b) Inconsistent member state responses | – Strong: Comprehensive tobacco bans, warning labels; – Enforcement variations | [1] Health AI high-risk (AI Act); [2] Contact-tracing (GDPR) | a. UK-EU co-ordination trouble (Brexit); b. Member state capacity variations |
| USA | i. Public Health Service Act, 1944; ii. HIPAA, 1996; iii. Controlled Substances Act, 1970 | a) Partial: CDC leads, but state autonomy caused inconsistent COVID-19 responses | – Partial: State-level tobacco laws, no federal FCTC ratification | [1] FDA regulates AI medical devices (21 CFR Part 860); [2] Predictive analytics for opioid crisis | a. Federal-state tensions; b. 8.6% uninsured (2023) |
| LMICs | i. Weak national health laws; ii. WHO/Africa CDC guidelines | a) Limited: 30% meet IHR core capacities; b) Delayed Ebola responses (2014–2016) | – Weak: 17% compliance with FCTC warning labels in Nigeria | [1] Nascent: Mobile-based surveillance in Kenya; [2] Limited by 29% internet penetration | a. Resource constraints; b. Donor-driven health budgets; c. Lack of AI/data privacy laws |

**Table 1: Comparative Analysis of International Public Health Instrument Implementation**

Table 1 delineates a comparative overview encapsulating the differential implementation of international public health instruments across the surveyed jurisdictions. This taxonomy synthesises salient legal frameworks, incorporates compliance gradations also including AI integration and challenges.

---

[93] *Ibid.*
[94] Nicaise Ndembi et al., "Integrating artificial intelligence into African health systems and emergency response: Need for an ethical framework and guidelines," 16 *Journal of Public Health in Africa* 876 (2025).
[95] Chinwe Iwu-Jaja et al., "New Vaccine Introductions in WHO African Region between 2000 and 2022," 11 *Vaccines* 1722 (2023).

## 4. Role of Artificial Intelligence in Public Health Governance

Artificial Intelligence constitutes a transformative vector within the domain of public health governance.[96] It augments the operationalisation of IPHIs[97]. In the taxonomy of disease surveillance, predictive models driven by Artificial Intelligence interrogate extensive data corpora. These datasets cover digital social platforms, travel registries, health reportage compendia, etc. This analytical process facilitates precocious detection of epidemiological outbreaks. These functionalities align with the prescriptive stipulations within Art. 5 of IHR mandating sophisticated surveillance capacities.[98] The Canadian Artificial Intelligence platform BlueDot exemplifies this technological evolution. It identified the early pattern of COVID-19 cases that began in Wuhan in Dec. 2019 which preceded the formal issuance of the PHEIC declaration by WHO.[99] The computational algorithms deployed by BlueDot operationalise ML[100], facilitating the processing of unstructured data matrices that extrapolate forthcoming disease dissemination trajectories, operationalising timely notification protocols mandated under IHR.[101] India integrated AI functionalities through the Aarogya Setu application for the purpose of contact tracing during the COVID-19 pandemic. This platform utilised Bluetooth signal mapping which integrated geospatial positioning system analytics and delineated vectors of pathogenic transmission. It fulfilled the requirements of Art. 6 of IHR concerning accelerated national response obligations.[102] EIOS platform of WHO deploys natural language processing methodologies, enabling the contemporaneous monitoring of global pathogenic threats. This fortifies alignment with surveillance imperatives codified within IHR.[103] Health data analytics constitutes a critical apparatus for the optimisation of resource distribution. This enhances formulation of public health policy architectures. The European Centre for Disease Prevention and Control operationalises artificial intelligence interfaces to synthesise methodologies which then interrogate epidemiological data corpora. They establish hierarchies of vaccine distribution exigencies. This process was exemplified during the COVID- 19 immunological dissemination campaign in 2021.[104] Within the United States, IBM Watson Health employs predictive analytic constructs forecasting the exigencies pertaining to hospital bed occupancy. This augments the public health responsiveness to epidemiological surges.[105] Telemedicine mechanisms and digital health infrastructures further operationalise AI deployments that are particularly monitored within diagnostic functionalities. The Artificial Intelligence model by the Google Health effectuates automated screening of retinal image for the identification of diabetic retinopathy with ninety percent diagnostic analyses precision. This model expands diagnostic accessibility within medically underserved areas corresponding to FCTC's emphasis upon the prevention of non-communicable disease proliferation.[106] LMICs exhibit embryonic integration of AI platforms. The mHealth initiative within Kenya exemplifies this developmental vector. This mobile-based platform facilitates pathogenic reporting though the structural infrastructural limitations attenuate scalability potentials within its territorial domains.[107] These technological applications evince the capacity of AI to reinforce the implementations of IHR as well as FCTC. It potentiates surveillance precision, enhancing allocative rationalisation in public health systems. They further recalibrate the indices of accessibility of health care services.

---

[96] Jennifer K Wagner, Megan Doerr and Cason D Schmit, "AI Governance: A Challenge for Public Health," 10 *JMIR Public Health and Surveillance* e58358–e58358 (2024).
[97] International Public Health Instruments.
[98] Giulio Bartolini, "The Failure of 'Core Capacities' Under the Who International Health Regulations," 70 *International and Comparative Law Quarterly* 233–50 (2021).
[99] Isaac I Bogoch et al., "Pneumonia of unknown aetiology in Wuhan, China: potential for international spread via commercial air travel," 27 *Journal of Travel Medicine* taaa008 (2020).
[100] Machine Learning.
[101] Abhishek Tripathi and Rachna Rathore, "AI in Disease Surveillance — An Overview of How AI Can Be Used in Disease Surveillance and Outbreak Detection in Real-World Scenarios," 1st ed., in R. Singh, A. Gehlot, *et al.* (eds.), *AI in Disease Detection* 337–59 (Wiley, 2025).
[102] Saurav Basu, "Effective Contact Tracing for COVID-19 Using Mobile Phones: An Ethical Analysis of the Mandatory Use of the Aarogya Setu Application in India," 30 *Cambridge Quarterly of Healthcare Ethics* 262–71 (2021).
[103] Foluke Ekundayo, "Using machine learning to predict disease outbreaks and enhance public health surveillance," 24 *World Journal of Advanced Research and Reviews* 794–811 (2024).

## 5. Conclusion

The confluence of jurisprudence and medical science, as embodied by IPHIs establishes a robust normative architecture for addressing transnational health issues. These articulate a harmonized legal armature that advances the domain. Comparative analysis across various jurisdictions reveals both paramount practices and persistent gaps in implementations. This relative analysis consists of India's constitutionally driven framework under Art. 21 and EDA 1897, EU's supranational coordination via Cross-Border Healthcare Directive and GDPR, US's decentralized governance under the PHS Act and CSA, followed by the resource constrained realities of LMICs. India's efforts highlights its innovative AI integration but challenges like rural-urban disparities subsists. EU's harmonized approach demonstrates institutional strength yet faces internal dissonances, while US's decentralized system fosters innovation but struggles with regulatory inconsistency. LMICs grapple with infrastructural scarcities and limited compliance with IHR and FCTC. This is despite emergent AI mechanisms like Kenya's mHealth. It enhances compliance with IHR through predictive surveillance (e.g., BlueDot's COVID- 19 detection), optimizing resource allocation (e.g., ECDC's vaccine prioritization), and expanding individual access (e.g., Google Health's retinopathy screening). However, issues with regard to data privacy, algorithmic bias and technological inconsistencies between high income countries and LMICs call for careful governance. EU's AI Act and GDPR give a model for regulating AI in health while LMICs need targeted technical assistance to tie digital gaps. Emerging technologies as blockchain for secure health data supervision and IoT for real- time monitoring, offer further potential to align technological innovation with global health law. Subsequently, strengthened transnational cooperation holds vitality. Global community can fortify public health systems by integrating AI and emerging technologies responsibly. The following recommendations further help advance scholarship, effectively confronting the complications of international health threats:

   i.   Strengthen WHO's Enforcement Mechanism
  ii.   Codify Global Algorithmic Governance
 iii.   Institutionalise AI Regulation in Health Governance
  iv.   Expand Technocratic and Juridical Capacitation in LMICs
   v.   Foster Intergovernmental Collaboration
  vi.   Operationalise Cross-Border Data Sharing Platforms
 vii.   Encourage Indo-EU Synergy in AI Public Health Innovation
viii.   Apply Interdisciplinary Doctrinal Models

---

[104] Adam J Kucharski, Emma B Hodcroft and Moritz U G Kraemer, "Sharing, synthesis and sustainability of data analysis for epidemic preparedness in Europe," 9 *The Lancet Regional Health - Europe* 100215 (2021).
[105] Altmann-Richer Lisa, "Using predictive analytics to improve health care demand forecasting" *Institute and Faculty of Actuaries* 19 (2018).
[106] Luis Filipe Nakayama et al., "Artificial intelligence for telemedicine diabetic retinopathy screening: a review," 55 *Annals of Medicine* 2258149 (2023).
[107] Don Lawrence Mudzengi et al., "Using mHealth Technologies for Case Finding in Tuberculosis and Other Infectious Diseases in Africa: Systematic Review," 12 *JMIR mHealth and uHealth* e53211 (2024).